\documentclass[aps,prb,superscriptaddress,reprint]{revtex4-1}
\usepackage[english]{babel}
\usepackage{amsmath,amsthm,amssymb}
\usepackage{amsfonts}
\usepackage{xspace}
\usepackage{bm}
\usepackage{graphicx}

\usepackage[separate-uncertainty = true, exponent-product = \times]{siunitx}
\usepackage{lmodern}
\usepackage{hyperref}
\usepackage{dcolumn}

\usepackage{color}
\newcommand{\red}[1]{#1}
\begin{document}
\newcommand{\la}{\langle}
\newcommand{\ra}{\rangle}
\newcommand{\eq}[1]{\begin{equation}#1\end{equation}}
\newcommand{\multeq}[1]{\begin{equation}\begin{split}#1\end{split}\end{equation}}
\newcommand{\p}{\partial}
\newcommand{\pfrac}[2]{\frac{\p#1}{\p#2}}
\renewcommand{\dfrac}[2]{\frac{d#1}{d#2}}
\newcommand{\mat}[1]{\left(\begin{smallmatrix}#1\end{smallmatrix}\right)}

\newcommand{\dT}{\Delta T_\mathrm{me}}
\newcommand{\gsm}{g_\mathrm{r}}

\renewcommand{\figurename}{FIG.}
\renewcommand{\tablename}{TABLE}

\title{Current heating induced spin Seebeck effect}%

\author{Michael Schreier}
\email{michael.schreier@wmi.badw.de}
\affiliation{Walther-Mei\ss ner-Institut, Bayerische Akademie der Wissenschaften, Garching, Germany}

\author{Niklas Roschewsky}
\affiliation{Walther-Mei\ss ner-Institut, Bayerische Akademie der Wissenschaften, Garching, Germany}

\author{Erich Dobler}
\affiliation{Walther-Mei\ss ner-Institut, Bayerische Akademie der Wissenschaften, Garching, Germany}

\author{Sibylle Meyer}
\affiliation{Walther-Mei\ss ner-Institut, Bayerische Akademie der Wissenschaften, Garching, Germany}

\author{Hans Huebl}
\affiliation{Walther-Mei\ss ner-Institut, Bayerische Akademie der Wissenschaften, Garching, Germany}

\author{Rudolf Gross}
\affiliation{Walther-Mei\ss ner-Institut, Bayerische Akademie der Wissenschaften, Garching, Germany}
\affiliation{Physik-Department, Technische Universit\"at M\"unchen, Garching, Germany}

\author{Sebastian~T.~B. Goennenwein}
\affiliation{Walther-Mei\ss ner-Institut, Bayerische Akademie der Wissenschaften, Garching, Germany}

\begin{abstract}
A measurement technique for the spin Seebeck effect is presented, wherein the normal metal layer used for its detection is exploited simultaneously as a resistive heater and thermometer. We show how the various contributions to the measured total signal can be disentangled, allowing to extract the voltage signal solely caused by the spin Seebeck effect. To this end we performed measurements as a function of the external magnetic field strength and its orientation. We find that the effect scales linearly with the induced rise in temperature, as expected for the spin Seebeck effect. 
\end{abstract}

\maketitle

The spin Seebeck effect~\cite{Uchida2008, Uchida2010, Uchida2010a, Uchida2010b} (SSE) is one of the hot topics in spin caloritronics.~\cite{Bauer2012} In analogy to the charge Seebeck effect, where a charge current is driven by an applied temperature gradient, in the spin Seebeck effect a spin current is driven by a temperature gradient.~\cite{Xiao2010} Since there is no direct meter for spin currents in present experiments usually a ferromagnet/normal metal (F/N) bilayer structure is used to convert the spin current into an electric signal: A temperature gradient applied perpendicular to the F/N bilayer drives a spin current across the F/N interface. This spin current is then converted into a charge current in N by virtue of the inverse spin Hall effect. Since most spin Seebeck effect measurements are performed using open circuit boundary conditions, the experimental signature of the spin Seebeck effect is a spin Hall electric field - viz. the corresponding spin Seebeck voltage - which is oriented perpendicular to both the applied temperature gradient and the magnetization in F.\\
Nowadays most spin Seebeck experiments are performed in the so-called longitudinal geometry, in which the temperature gradient and the spin current are parallel, and oriented perpendicular to the F/N interface. To rule out anomalous Nernst effect voltages in F,~\cite{Huang2011,Huang2012} this geometry however requires that the (ferro- or ferri-) magnetic constituent is insulating. In most longitudinal spin Seebeck experiments to date, the magnetic insulator yttrium iron garnet (Y$_3$Fe$_5$O$_{12}$, YIG) is used for F, and the high Z metals Pt or Au are used for N.~\cite{Uchida2008, Uchida2010, Uchida2010a, Uchida2010b, Weiler2012, Qu2013}\\
In experiments, the controlled generation and quantification of temperature gradients represents a challenge. The temperature gradients are most often established by clamping the F/N sample between two heat reservoirs, acting as heat source and sink.~\cite{Uchida2008, Uchida2010, Uchida2010a, Uchida2010b, Jaworski2010, Qu2013, Meier2013, Kikkawa2013} An important issue in this type of spin Seebeck effect setup is good thermal coupling between the heat reservoirs and the sample. Laser heating~\cite{Weiler2012} is an alternative technique, which enables scannable, local temperature gradient generation. The temperature gradients thus generated, however, can be quantified only from numerical temperature profile calculations.~\cite{Schreier2013}\\
\begin{figure}%
\includegraphics[width=\columnwidth]{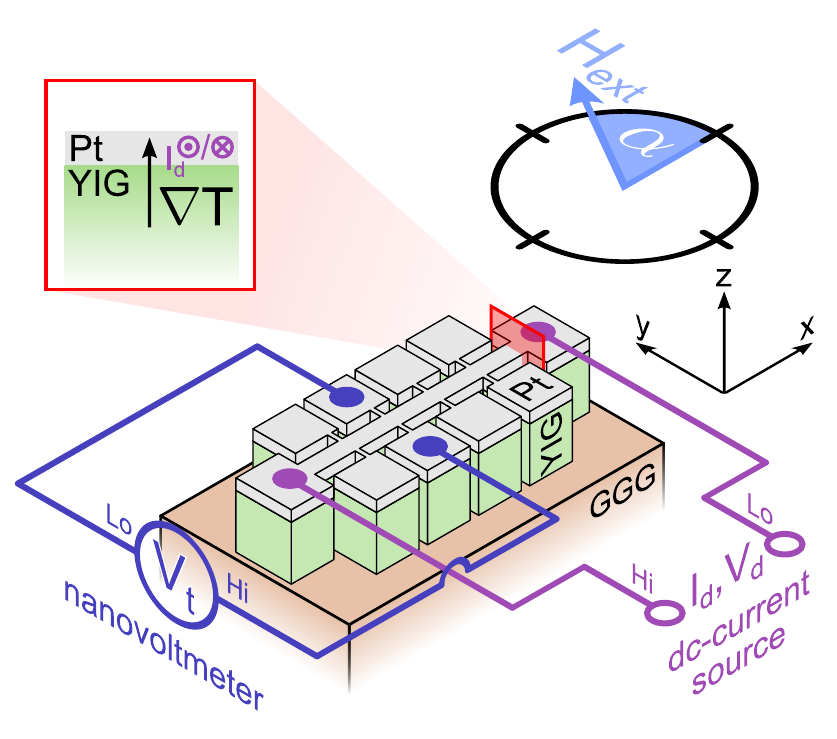}%
\caption{Sketch of the setup used for the current heating induced spin Seebeck experiments. The samples consists of magnetic insulator (YIG) thin films on single crystalline GGG or YAG substrates covered by a thin normal metal (Pt) film. The YIG/Pt bilayer is patterned into a Hall bar mesa structure. A dc-current source is used to drive a large current $I_\mathrm{d}$ through the Hall bar while the voltage drop $V_\mathrm{t}$ transverse to the current direction is measured with a nanovoltmeter. An external, in-plane, magnetic field is applied at an angle $\alpha$ to the current direction. Due to the resistive (Joule) heating by $I_\mathrm{d}$ a temperature gradient across the F/N interface emerges, giving rise to the spin Seebeck effect.}%
\label{fig:setup}%
\end{figure} 
In this paper we present a third, very simple technique to generate large thermal gradients across the F/N interface. The main idea is to use the sample's normal metal layer itself as a resistive heater. In other words, we drive a large dc-current $I_\mathrm{d}$ through N, and simultaneously record the thermal (spin Seebeck) voltage in the direction transverse to the driving current (Fig.~\ref{fig:setup}).\\

Since the current heating induced spin Seebeck voltage $V_\mathrm{iSSE}$ originates form the inverse spin Hall effect, we expect $V_\mathrm{iSSE}\propto \pmb j_\mathrm{s}\times\hat{\pmb s}$, where $\pmb j_\mathrm{s}$ is the direction of the spin current and $\hat{\pmb s}$ is its polarization vector. This can be used to discriminate $V_\mathrm{iSSE}$ from other possible signal contributions. For $H_\mathrm{ext}$ in the sample plane along $I_\mathrm{d}$ (along $x$, $\alpha=0^\circ$), a spin Seebeck voltage will arise along $y$. The large voltage drop $V_\mathrm{d}$ arising along the direction of current flow thus will not influence the spin Seebeck measurement.\\
The spin Seebeck effect is, in fact, driven by the temperature difference $\dT$ between magnons in F and electrons in N rather than the thermal gradient across the layers.~\cite{Xiao2010} While the precise determination of $\dT$ is very challenging~\cite{Schreier2013} and depends crucially on e.g. sample dimension and the material parameters at the chosen sample temperature, $\dT$ will, in first order approximation, be directly proportional to the temperature increase (decrease) of N with respect to the heat sink in the experiment (i.e. it is proportional to the the thermal gradient).~\cite{Xiao2010, Schreier2013} This temperature increase is in turn directly proportional to the dissipated electrical power (the Joule heating) $P_\mathrm{Joule}=V_\mathrm{d}I_\mathrm{d}=I_\mathrm{d}^2R$ where $R$ is the sample resistance. Using an insulating ferromagnet (YIG) greatly simplifies the interpretation of the experimental results since the current will only flow in the normal metal (platinum). \red{The heat in our experiment is generated uniformly within the entire N layer as compared to an injection through the top interface only for the clamping technique or the nonuniform heating for the laser method. Nevertheless, for a fixed amount of heat, in steady state and since the spin Seebeck effect is generated at the F/N interface rather than within N, the thermal gradient at the F/N interface should be very similar among the techniques.} In summary, we thus expect
\begin{equation}
	V_\mathrm{iSSE}\propto I_\mathrm{d}^2\cos\alpha,
\label{eq:Vsseprop}
\end{equation}
However, the voltage \red{$V_\mathrm{t}=E_\mathrm{t}\times w$ ($E_\mathrm{t}$ and $w$ being the transverse electric field under open circuit conditions and the width of the Hall bar, respectively)} transverse to $I_\mathrm{d}$ will have contributions from the spin Seebeck effect and from magnetoresistive effects, such as the newly discovered spin Hall magnetoresistance.~\cite{Nakayama2013,Althammer2013a,Chen2013a} Typically, these magnetoresistive transverse voltages will be much larger than the $V_\mathrm{iSSE}$ of interest. 
Additionally, the longitudinal resistance can contribute to $V_\mathrm{t}$ due to a slight misalignment of the transverse contacts. Since these effects are linear, or odd in $I_\mathrm{d}$, they can be distinguished from thermal effects, proportional to $P_\mathrm{Joule}$ or $I_\mathrm{d}^2$, by comparing two measurements with reversed driving current direction. 
The resistive contributions and the cross-coupling obey $V_\mathrm{res}(+I_\mathrm{d})=-V_\mathrm{res}(-I_\mathrm{d})$ while the spin Seebeck voltage obeys $V_\mathrm{iSSE}(+I_\mathrm{d})=+V_\mathrm{iSSE}(-I_\mathrm{d})$. $V_\mathrm{iSSE}$ can thus be obtained by adding \red{$V_\mathrm{t}(+I_\mathrm{d})$ to $V_\mathrm{t}(-I_\mathrm{d})$} such that
\begin{eqnarray}
	V_\mathrm{t}(+I_\mathrm{d}) + V_\mathrm{t}(-I_\mathrm{d}) &=& V_\mathrm{res}(+I_\mathrm{d}) + V_\mathrm{res}(-I_\mathrm{d}) +\notag\\
	&& V_\mathrm{iSSE}(+I_\mathrm{d}) + V_\mathrm{iSSE}(-I_\mathrm{d})\notag\\
	&=& V_\mathrm{res}(+I_\mathrm{d}) - V_\mathrm{res}(+I_\mathrm{d}) +\notag\\
	&& V_\mathrm{iSSE}(+I_\mathrm{d}) + V_\mathrm{iSSE}(+I_\mathrm{d})\notag\\
	&=& 2V_\mathrm{iSSE}(+I_\mathrm{d})
\label{eq:VpmI}.
\end{eqnarray}
It is fair to argue that with increasing $I_\mathrm{d}$, the sample's resistance $R=R(T)=R(I_\mathrm{d}^2)$ will increase due to the induced temperature changes. This also influences the resistive contributions by introducing higher order terms which, in good approximation, should be odd powers of $I_\mathrm{d}$ since 
\begin{equation}
	V_\mathrm{res}\propto I_\mathrm{d}\red{\times} R\propto I_\mathrm{d}\red{\times} I_\mathrm{d}^2.
\label{eq:Vresodd}
\end{equation}
Thus they should cancel out in the aforementioned procedure.\\
\red{We would also like to add that it is possible to extract $V_\mathrm{iSSE}$ from the longitudinal voltage as well, albeit generally with a worse signal to noise ratio due to the large background signal ($V_\mathrm{d}$).}\\

The samples in our experiment consists of YIG thin films grown by pulsed laser deposition on $\SI{500}{\micro m}$ thick gadolinium gallium garnet (Gd$_3$Ga$_5$O$_{12}$, GGG) or yttrium aluminium garnet (Y$_3$Al$_5$O$_{12}$, YAG) substrates. On top of the YIG layer few $\SI{}{\nano m}$ thick Pt films were then deposited in situ, without breaking the vacuum, using electron beam evaporation (more details on the sample growth can be found in Refs.~\onlinecite{Geprags2012,Althammer2013a}). One sample was fabricated with an additional gold spacer layer between YIG and Pt. After removing the samples from the growth chamber, the Pt (Au) and the YIG were patterned into Hall bar mesa structures (length $\SI{950}{\micro m}$, width $\SI{80}{\micro m}$) using optical lithography and argon ion beam milling. Afterwards the samples are mounted onto copper heat sinks.\\ % and mounted into a custom built sample carrier system.\\
The measurements in this paper have all been performed in vacuum ($p\lesssim\SI{1}{\milli\bar}$) in a cryostat with variable temperature insert, with the sample stabilized at a base temperature of $\SI{250}{K}$. Note, however, that measurements under ambient conditions in an electromagnet at room temperature (not shown) gave very similar results. We furthermore measured the temperature dependence of $\rho_\mathrm{Pt}$ by systematically changing the cryostat base temperature. In this way, the Pt resistance can be used for on-chip thermometry~\cite{Ri1991, Ri1993} in the subsequent experiments.\\
\begin{figure}%
\includegraphics[width=\columnwidth]{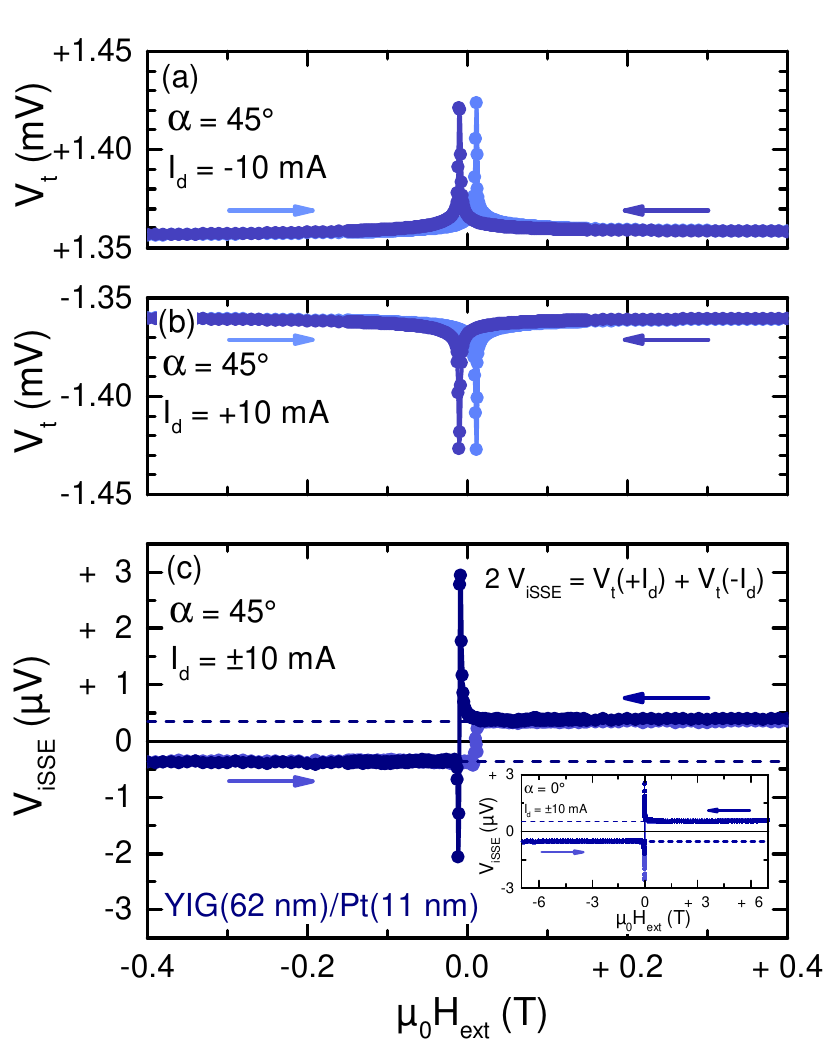}%
\caption{Recorded transverse voltage on the GGG/YIG($\SI{61}{\nano m}$)/Pt($\SI{11}{\nano m}$) sample as a function of the external magnetic field strength for $\alpha=45^\circ$. \red{The arrows indicate the sweep direction of the external magnetic field in the experiment.} 
\textbf{(a)} For $I_\mathrm{d}<0$ a positive offset voltage signal is recorded that exhibits the typical features of the spin Hall magnetoresistance. \textbf{(b)} Reversing the direction of $I_\mathrm{d}$ also inverts the observed voltage signal. \textbf{(c)} Adding $V_\mathrm{t}(+I_\mathrm{d})$ and $V_\mathrm{t}(-I_\mathrm{d})$ %and dividing the result by two 
reveals the much smaller, thermal (spin Seebeck) component. The large spikes close to the YIG's coercive fields are attributed to either domain reconfiguration or spin torque effects. The inset shows $V_\mathrm{iSSE}$ at large fields for $\alpha=0^\circ$. Here $V_\mathrm{iSSE}$ stays constant for fields of up to $\SI{7}{T}$.
}%
\label{fig:2}%
\end{figure}
The $V_\mathrm{iSSE}$ extraction procedure is visualized in Fig.~\ref{fig:2} for a fixed angle $\alpha=45^\circ$ between the Hall bar and the external magnetic field (\textit{cf.} Fig.~\ref{fig:setup}) on a GGG/YIG($\SI{61}{\nano m}$)/Pt($\SI{11}{\nano m}$) sample. Here the transverse voltage is recorded as a function of the external magnetic field magnitude, which is varied from $+\SI{0.4}{T}$ to $-\SI{0.4}{T}$ and back to $+\SI{0.4}{T}$. For a pure spin Seebeck signal one would expect the observed signal's shape to closely mimic that of the magnetic hysteresis loop of YIG, but apparently this is not the case. Clearly the signal shown in panel \textbf{(a)} is dominated by the transverse component of the in-plane spin Hall magnetoresistance,~\cite{Althammer2013a} which changes sign upon changing the current direction [panel \textbf{(b)}]. Upon adding the two curves, however, the hysteresis loop becomes visible [panel \textbf{(c)}]. \red{For $H_\mathrm{ext}>0$ we observe a  positive $V_\mathrm{iSSE}$ as observed in earlier experiments.~\cite{Uchida2010}}
%%%%
The large additional peaks close to the coercive fields may stem from torque induced magnetization dynamics~\cite{Fan2013} in the YIG, which affect the spin current flow in the Pt \red{or may be an artifact due to stray Oersted fields}. Dedicated experiments will be required to pinpoint the exact origin of these signal contributions. We can, however, exclude a proximity effect~\cite{Huang2011} induced origin since these peaks appear in \textit{all} samples, including the one with the additional gold layer between the YIG and the Pt. Here, we are interested only in the spin-Seebeck-like signal at high fields, which stays constant up to $\SI{7}{T}$ (inset Fig.~\ref{fig:2}).\\
\begin{figure}%
\includegraphics[width=\columnwidth]{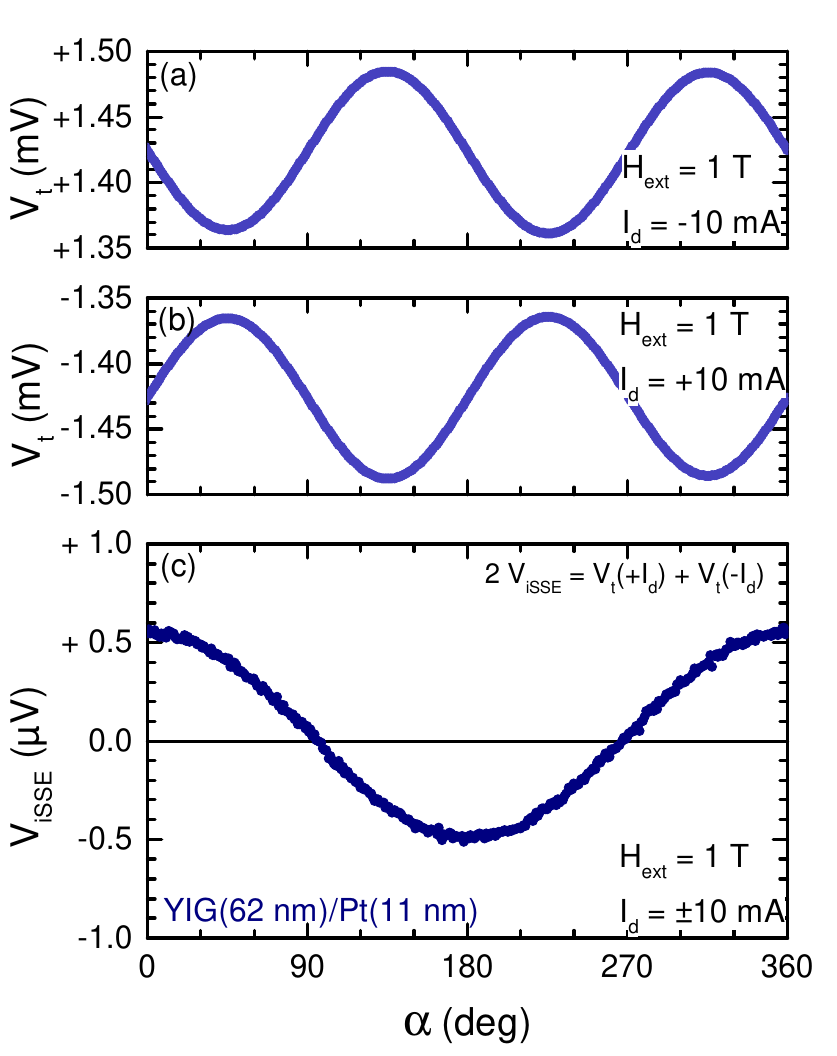}%
\caption{Thermal voltage as a function of the external magnetic field direction $\alpha$ on the GGG/YIG($\SI{61}{\nano m}$)/Pt($\SI{11}{\nano m}$) sample. The magnitude of the external magnetic field remains fixed at $\SI{1}{T}$ throughout the entire measurement. \textbf{(a)} For $I_\mathrm{d}<0$ a positive offset voltage is recorded with the visible $\sin^2\alpha$ variation stemming from the spin Hall magnetoresistance. \textbf{(b)} Inverting the current ($I_\mathrm{d}>0$) also reverses the voltage signal, but upon adding up $V_\mathrm{t}(+I_\mathrm{d})$ and $V_\mathrm{t}(-I_\mathrm{d})$ and dividing the result by two [\textbf{(c)}] a $\cos\alpha$ component remains, consistent with the spin Seebeck effect [Eq.~\eqref{eq:Vsseprop}].}%
\label{fig:V(alpha)}%
\end{figure} 
To investigate the expected $\cos\alpha$ dependence of $V_\mathrm{iSSE}$ we keep the applied magnetic field at a fixed value of $\SI{1}{T}$ and record the transverse voltage $V_\mathrm{t}$ while varying the field orientation with respect to the Hall bar. The field value is chosen large enough to rule out any remanent magnetic features of the YIG and ensure that its magnetization is truly parallel to the external magnetic field. Figure~\ref{fig:V(alpha)} shows the result of this measurement on the GGG/YIG($\SI{61}{\nano m}$)/Pt($\SI{11}{\nano m}$) sample. As in Fig.~\ref{fig:2} the measured signal is dominated by the spin Hall magnetoresistance, featuring its characteristic $\sin^2\alpha$ dependence, which reverses sign as the current direction is inverted. Once again, by adding the signals obtained for opposite current direction the resistive effects cancel out and the spin Seebeck component remains. As predicted by Eq.~\eqref{eq:Vsseprop} the signal follows a $\cos\alpha$ dependence.\\
\begin{figure}%
\includegraphics[width=\columnwidth]{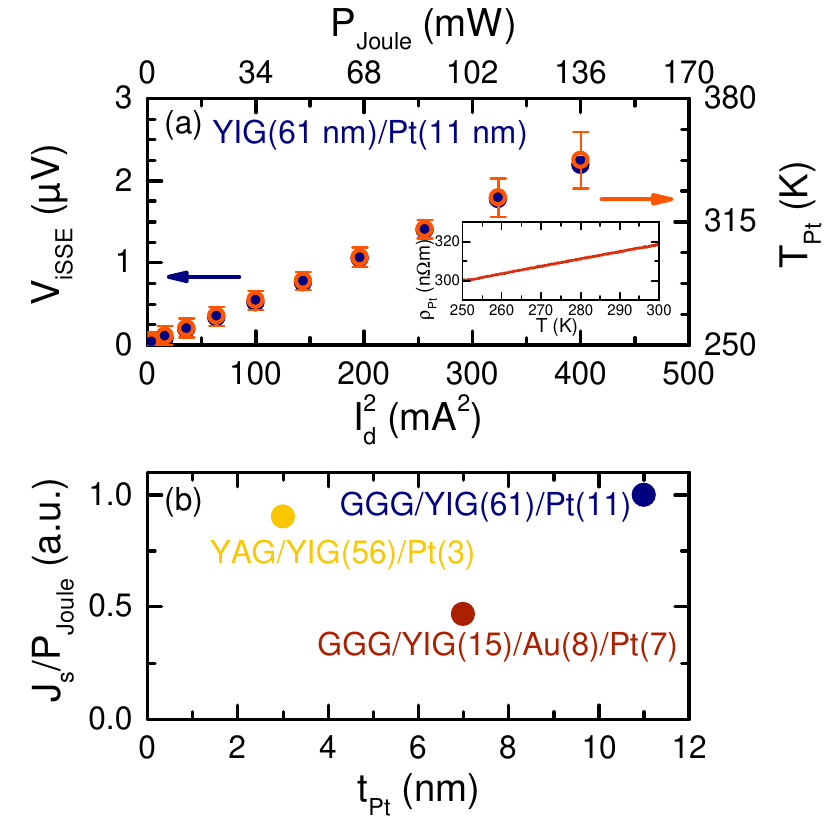}%
\caption{\textbf{(a)} Thermal component ($V_\mathrm{iSSE}$) of the recorded voltage (full blue symbols) and Pt temperature (open circles) as a function of the square of applied current on the GGG/YIG($\SI{61}{\nano m}$)/Pt($\SI{11}{\nano m}$) sample. To obtain the individual $V_\mathrm{iSSE}(I_\mathrm{d})$ data points the external magnetic field was rotated at a fixed field strength of $\SI{1}{T}$, from which the spin Seebeck voltage is extracted [$V_\mathrm{iSSE}\equiv V_\mathrm{iSSE}(\alpha=0^\circ)$] for each value of the driving current $\pm I_\mathrm{d}$. The observed $V_\mathrm{iSSE}$ scales quadratically with $I_\mathrm{d}$ as does $T_\mathrm{Pt}$. \red{The indicated error for $T_\mathrm{Pt}$ is an upper estimate for the uncertainty in the fitting algorithm and accounts for the fact that the two rightmost data points were extrapolated. The inset shows the resistivity of the Pt film as a function of temperature.} \textbf{(b)} Generated spin current per applied heating power for the samples investigated in this letter. The numbers in parentheses give the layer thicknesses in $\SI{}{\nano m}$. Taking the smaller spin mixing conductance of the YIG/Au interface into account, all samples give very similar spin current generation efficiencies.}%
\label{fig:V(I)}%
\end{figure}
To confirm that the recorded signals indeed stem from a thermal effect this procedure is repeated as a function of the applied current. Panel \textbf{(a)} in Fig.~\ref{fig:V(I)} shows $V_\mathrm{iSSE}=\frac12[V_\mathrm{t}(+I_\mathrm{d})+V_\mathrm{t}(-I_\mathrm{d})]$ as a function of $I_\mathrm{d}^2$ on the GGG/YIG($\SI{61}{\nano m}$)/Pt($\SI{11}{\nano m}$) sample. $V_\mathrm{iSSE}$ clearly shows a quadratic dependence on the applied current, supporting the notion of Eq.~\eqref{eq:Vsseprop} that the measured spin Seebeck effect should scale quadratically with $I_\mathrm{d}$. Moreover the effect quickly drops below the noise floor for small currents for which the spin Hall magnetoresistance is still clearly visible in $V_\mathrm{t}$.~\cite{Althammer2013a} Furthermore, by simultaneously measuring the resistance of the Pt Hall bar along the current direction we are also able to determine its temperature by comparing the measured resistance value to a $R(T)$ calibration curve recorded separately at a small current value \red{(inset Fig.\ref{fig:V(I)})}.  As expected, the temperature increase of the Pt film is directly proportional to $I_\mathrm{d}^2$ as well. In other words, $V_\mathrm{iSSE}$ is directly proportional to the temperature increase of the Pt film as suggested above. 
Panel \textbf{(b)} in Fig.~\ref{fig:V(I)} compares the results of the different samples investigated in this letter. 
Here the recorded spin Seebeck voltage is normalized as to extract the spin current density per applied heating power $J_\mathrm{s}/P_\mathrm{Joule}$. This is achieved by dividing $V_\mathrm{iSSE}$ by the sample resistance, Joule heating power and the correction factor for spin diffusion in the normal metal.~\cite{Schreier2013} The latter is calculated using a spin diffusion length of $\SI{1.5}{\nano m}$ for Pt~\cite{Althammer2013a} and assuming that the gold layer does not affect the spin current. 
The value of $J_\mathrm{s}/P_\mathrm{Joule}$ is very similar for the investigated YIG/Pt samples, while the GGG/YIG($\SI{15}{\nano m}$)/Au($\SI{8}{\nano m}$)/Pt($\SI{7}{\nano m}$) sample shows about half the value of the YIG/Pt samples, fully consistent with the smaller spin mixing conductance of the YIG/Au interface.~\cite{Althammer2013a, Weiler2013, Burrowes2012} \red{Generally samples with thinner YIG films give smaller voltage signals, the number of samples investigated here is, however, too small and the individual samples too different to confidently read any trend~\cite{Kehlberger2013} from this observation.}\\
The arguments brought forward here would also apply to a potential contamination via the anomalous Nernst effect. However, recent experiments~\cite{Geprags2012, Kikkawa2013, Geprags2013} show that pure Pt does not get proximity polarized by the YIG layer and hence no anomalous Nernst can occur. This is also confirmed by the measurement on the GGG/YIG($\SI{61}{\nano m}$)/Au($\SI{8}{\nano m}$)/Pt($\SI{7}{\nano m}$) sample which gives voltage signals very similar to those on the samples without the gold spacer layer [Fig.~\ref{fig:V(I)}\textbf{(c)}].\\

In summary, we have demonstrated that the longitudinal spin Seebeck effect can be measured by simply using the normal metal (Pt) layer as a Joule heater to create the required thermal gradient at the ferromagnet/normal metal interface. Measurements as a function of the magnitude and orientation of the applied magnetic field show the characteristic dependencies of the spin Seebeck effect and scale quadratically with the applied current, as expected for a thermal effect. We thus conclude that the simple Joule heating technique indeed enables the detection of the spin Seebeck effect in yttrium iron garnet/platinum thin film hybrid structures.\\
\\
We thank Timo Kuschel for valuable discussions and gratefully acknowledge financial support from the DFG via SPP 1538 ``Spin Caloric Transport'' (project GO 944/4-1) and the German Excellence Initiative via the Nanosystems Initiative Munich (NIM).

\bibliography{Current_heating_induced_spin_Seebeck_effect}
\end{document}